\documentclass[prc,twocolumn,amsmath,amssymb]{revtex4}

\usepackage{graphicx}
\usepackage{dcolumn}
\usepackage{bm}
\usepackage{amsmath}
\usepackage{float}

\begin{document}

\title{Competition between weak and $\alpha$-decay modes in superheavy nuclei}

\author{P.~Sarriguren}
\email{p.sarriguren@csic.es}
\affiliation{
  Instituto de Estructura de la Materia, IEM-CSIC, Serrano 123,
  E-28006 Madrid, Spain}


\begin{abstract}

The competition between $\alpha$- and $\beta^+/EC$-decay modes is studied 
systematically in nuclei along the $\alpha$-decay chains following the 
synthesis of superheavy nuclei with $Z=119$ and $Z=120$. A microscopic
approach based on deformed self-consistent Hartree-Fock mean-field
calculations with Skyrme forces and pairing correlations is used to describe
the $\beta^+$ and electron capture weak decays, whereas the $\alpha$-decay is 
estimated from existing phenomenological expressions. It is shown that 
$\alpha$-decay is in most cases the dominant decay mode, but interesting 
instances are identified where the half-lives are comparable, opening the 
possibility of new pathways towards more neutron-rich nuclei in this region.

\end{abstract}

\maketitle

\section{Introduction}

The study of superheavy nuclei (SHN), both theoretically and experimentally,
has been one of the most active research topics in nuclear physics during the
last decades. Comprehensive overviews of the experimental methods used to
synthesize SHN and theoretical approaches employed to their understanding can
be found in Refs. \cite{special_issue,hofmann_00,oganessian_07,hamilton_13,oganessian_15a,hofmann_16,giuliani_19,adamian_epja56}.

Despite the large number of SHN already synthesized with different types of
cold and hot fusion reactions, there is a limitation in these reactions that
produce neutron-deficient isotopes, far away from the predicted location of the
most stable SHN, which is unreachable by fusion reactions with stable beams.
Predictions of shell closures from macroscopic-microscopic models based on
deformed liquid drop models with shell corrections
\cite{patyk_91,moller_94,adamian_epja54,adamian_epja57} agree with predictions
from self-consistent non-relativistic \cite{rutz_97,kruppa_00,bender_01} and
relativistic \cite{meng_06,agbemava} mean-field models with effective nuclear
interactions that produce shell closures at $(Z,N)=(114,184),\ (120,172)$, and
$(126,184)$, depending on the effective interaction used.
The sensitivity of the shell closures to the properties of the underlying
nuclear forces, makes it possible to use SHN as a laboratory to investigate
the nuclear force. 

Because of the above mentioned experimental difficulties to reach these regions
of the nuclear chart, alternative methods are being explored to synthesize more 
neutron-rich SHN. These methods include fusion reactions with radioactive ion 
beams and multinucleon transfer reactions \cite{adamian_epja56,zagrebaev}, as 
well as fusion-evaporation reactions where charged particles are emitted from 
the compound nucleus \cite{lopez_19,hessberger_19,hong,adamian_varios}.

Knowledge of the decay properties of SHN is of paramount importance for their
identification and for understanding their nuclear structure. The planning,
execution, and analysis of experiments leading to SHN production requires a
detailed knowledge of the decay modes and half-lives of nuclei in a very wide
range of neutron and proton numbers. Nuclei synthesized in such fusion reactions
are basically identified by studying their decay modes, which essentially are
dominated by $\alpha$ decay and spontaneous fission (SF). Nevertheless, the
study of weak decays is also important, not only to fully characterize all the
possible decay modes, but also as a pathway thorough the predicted island of
stability. Weak decays, including electron captures (EC), are not usually
considered, partly due to the experimental difficulty to investigate them at
the low production rates achieved in present experiments. However, new
experimental techniques involving measurements of delayed coincidences between
X-rays from the EC process and SF or $\alpha$ decay of the daughter nucleus 
have been applied to the cases of $^{257}$Rf \cite{hess_257rf}, $^{258}$Db 
\cite{hess_258db}, and $^{244}$Md \cite{khuyagbaatar} and their EC decays have 
been determined.

Theoretically, $\beta^+/EC$ decays are also more difficult to study. The nuclear
structures of parent and daughter nuclei are involved in the process and thus
require a microscopic approach to evaluate the nuclear matrix elements
connecting the initial with all the final states reached in the process.
Present theoretical predictions of weak-decay half-lives in SHN are based on
different approaches. Purely phenomenological parametrizations have been used
\cite{zhang_06} to extrapolate the half-lives to SHN regions, where they are
unknown. Other type of calculations neglecting nuclear structure effects as well,
can be found in Refs. \cite{fiset_72,karpov_12,singh_20}. Those approaches 
consider only transitions connecting the ground states of parent and daughter 
nuclei, whereas their nuclear matrix elements are assumed to be constant with 
phenomenological values determined globally for all nuclei, thus neglecting
any structural effect. However, the values of such matrix elements may change by
orders of magnitude in different calculations. $Q_{EC}$ energies are taken from
phenomenological mass models. Therefore, the final estimation of the weak-decay
half-lives in those models is reduced to the calculation of the phase-space
factors. Half-lives for $\beta^+/EC$ decay were also calculated within a 
proton-neutron quasiparticle random-phase approximation (pnQRPA) that starts with 
a phenomenological folded-Yukawa single-particle Hamiltonian \cite{moller_19}, 
using masses from a droplet model and standard phase factors. 

In this work, $\beta^+/EC$-decay half-lives of SHN are studied from a 
microscopic point of view, following the work already started in Refs. 
\cite{sarri_shn_prc,sarri_shn_jpg,sarri_shn_plb}. The method is based on the 
pnQRPA approach where the parent and daughter partners involved in the decay 
are described microscopically from a deformed self-consistent Hartree-Fock (HF) 
calculation with Skyrme interactions and pairing correlations in the 
Bardeen-Cooper-Schrieffer (BCS) approximation (HF+BCS). In addition to the
cases of special interest studied in Refs.
\cite{sarri_shn_prc,sarri_shn_jpg,sarri_shn_plb}, a systematic study of $\alpha$
and $\beta^+/EC$ decay in the SHN is carried out in this work. Rather than the
study of decays in isotopic chains, the focus here is the competition between
decay modes in the $\alpha$-decay chain members that follow the synthesis of
SHN, including nuclei that are already reachable or are potentially accessible
with current (or near future) technological capabilities.

The identification of mass regions where $\alpha$ and $\beta^+/EC$ decay may 
compete, would guide experimental work aimed at finding those weak decay modes. 
Driven by this purpose, different $\alpha$ decay chains, characterized by 
fixed numbers of $N-Z$, are systematically studied starting with the new 
elements to be produced at $Z=119,120$ and ending at the nuclides with 
negative $Q_{EC}$ energies, where no further $\beta^+/EC$ decay is possible.

The paper is organized as follows. Section II contains a brief summary of the
theoretical formalism used to calculate the energy distribution of the 
Gamow-Teller (GT) strength, as well as the $\beta^+/EC$-decay half-lives. 
Section III contains the $\alpha$ and $\beta^+/EC$-decay half-lives for the 
SHN involved in the different $\alpha$-decay chains considered in this work, 
discussing the competition between those decay modes. Section IV contains 
the final remarks and conclusions.

\section{Theoretical formalism for weak decays}

The theoretical formalism used in this work to obtain the $\beta^+/EC$-decay
half-lives follows the lines described in Refs.
\cite{sarri_shn_prc,sarri_shn_jpg,sarri_shn_plb} for SHN. More details of
the formalism, as well as the various sensitivities of the half-lives to the
model ingredients can be found in Refs. \cite{sarri1,sarri2,sarri3,sarri_odd}.

The $\beta^+/EC$-decay half-life, $T_{\beta^+/EC}$, is obtained after summing all
the GT strengths $B(GT,E_{ex})$ to states with excitation energies $E_{ex}$ in 
the daughter nucleus lying below the $Q_i$ energy ($i=\beta^+,EC$).
The strength is weighted with phase-space factors $f^i(Z,Q_i-E_{ex})$,

\begin{equation}
T_i^{-1}=\frac{\left( g_{A}/g_{V}\right) _{\rm eff} ^{2}}{D}
\sum_{E_{ex} < Q_{i}}f^i \left( Z,Q_i-E_{ex} \right) B(GT,E_{ex}) \, ,
 \label{t12}
\end{equation}
with $D=6143$~s. A quenching factor of $0.77$ is used, such that 
$(g_A/g_V)_{\rm eff}=0.77(g_A/g_V)_{\rm free}$, with 
$(g_A/g_V)_{\rm free}=-1.270$. The total
half-life for the combined process is given by
$ T_{\beta^+/EC}^{-1} = T_{\beta^+}^{-1} + T_{EC}^{-1} $.

The nuclear structure calculation used to obtain the GT strength distribution
starts with a self-consistent calculation of the mean field.
This is achieved from an axially deformed HF calculation with Skyrme interactions
and pairing correlations in the BCS approximation using phenomenological gap
parameters. The Skyrme force SLy4 \cite{chabanat} is chosen for that purpose due to 
its capability to  account for a wide variety of nuclear properties 
throughout the nuclear chart. The mean field calculation generates  wave functions,
single-particle energies, and occupation amplitudes. The formalism used to solve
the HF equations was developed in Ref. \cite{vautherin}, assuming time reversal
and axial symmetry. The eigenstates of an axially deformed harmonic oscillator 
potential expressed in cylindrical coordinates are used to expand the 
single-particle wave functions, using 16 major shells.
This basis size is large enough to get convergence of the HF energy. It should 
also be noted that the use of such a basis, which is adjusted for each nucleus 
and interaction in terms of the oscillator length and axis-ratio parameters, 
accelerates the convergence of the results as compared with the spherical basis.

In the mean-field approach, the energy of the different shape configurations can
be evaluated with constrained calculations, minimizing the HF energy under the
constraint of keeping fixed the nuclear quadrupole deformation parameter $\beta_2$.
By varying $\beta_2$ one obtains deformation-energy curves (DECs),
where the various minima correspond to equilibrium nuclear shapes with the
absolute minimum being the ground state.
Deformation has been shown to be a critical ingredient in understanding the decay
properties of $\beta$-unstable nuclei \cite{sarri1,sarri2,sarri3,sarri_odd} and
this is also expected in SHN.
DECs of some representative cases of SHN, namely,
$^{252}$Fm, $^{266}$Sg, $^{290}$Fl, $^{294}$Lv, and $^{300}$120, were depicted
in Refs. \cite{sarri_shn_prc,sarri_shn_jpg,sarri_shn_plb}.
Their analysis shows that the ground states appear at prolate
deformations around $\beta_2 \approx 0.3$ in most of the SHN studied here. Only the 
heaviest nuclei with $Z > 116$ or $N > 170$ exhibit absolute energy minima
located around the spherical configuration within a flat region between
$\beta_2=-0.1$ and $\beta_2=0.1$. In this work we only consider half-lives
from the ground state of the parent nuclei.
Half-lives corresponding to shape configurations different from the ground
state were studied in Refs. \cite{sarri_shn_prc,sarri_shn_jpg,sarri_shn_plb}.

The nuclear structure of odd-$A$ nuclei is described within the
equal filling approximation (EFA), blocking the state of the unpaired nucleon
characterized by a given spin and parity. In EFA half of the odd
nucleon is placed in a given orbital, while the other half is placed in its 
time-reversed partner.
The EFA prescription represents a significant numerical advantage because
time-reversal invariance is preserved. By comparing EFA results
with those from more sophisticated approaches, it has been shown
\cite{schunck_10} that EFA is very reliable and precise for most practical
applications. 

In cases where the spin and parity, $J^{\pi}$, of the nucleus is experimentally
determined, the natural choice is to select the blocked state according to the
experimental assignments, although in many cases they are derived from systematics.
As expected, the experimental assignments are found within the calculated states 
in the neighborhood of the Fermi level in all the cases studied here.
When there is no experimental information on $J^{\pi}$, the
blocked state is chosen as the state that minimizes the energy.
The sensitivity of the half-lives to $J^{\pi}$ has been studied
in Refs. \cite{sarri_shn_prc,sarri_shn_jpg,sarri_shn_plb}, 
where calculations were performed for several states close to the Fermi energy
with opposite parity.
This study is especially interesting given that small changes in the theoretical 
description of the nucleus can lead to different $J^{\pi}$ values for the ground 
states, which in turn determine to a large extent the spin and parity of the
states reached in the daughter nucleus because of the selection rules of the
GT operator.

\begin{figure}[th]
\includegraphics[width=85mm]{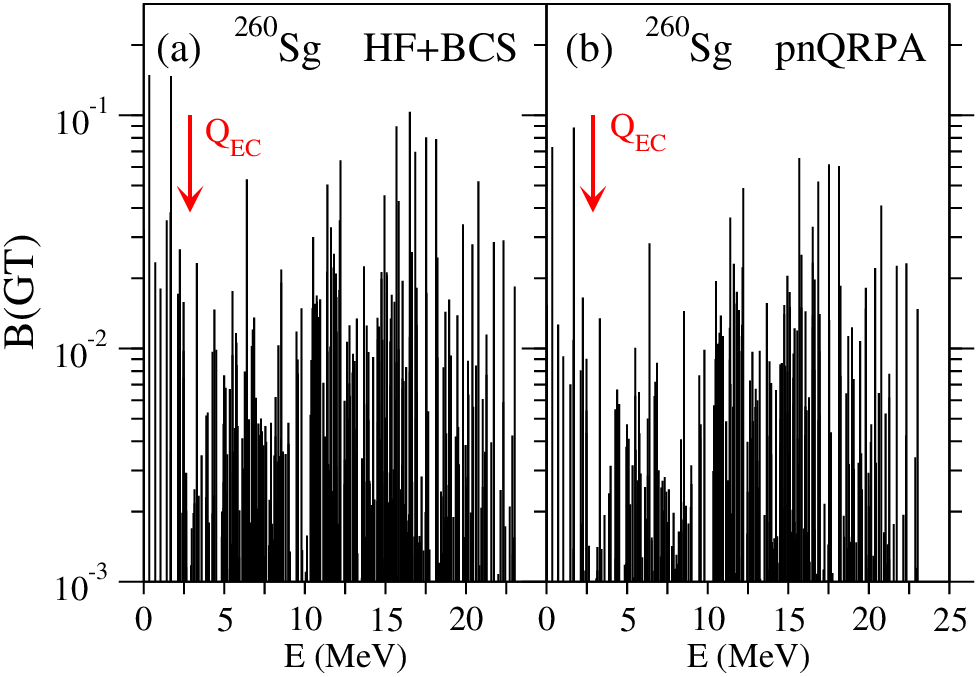}
\caption{GT strength distribution of $^{260}$Sg calculated within (a) HF+BCS(SLy4) 
and (b) pnQRPA with spin-isospin separable forces.}
\label{fig_bgt}
\end{figure}

Once the mean field is constructed, the nuclear matrix elements of the Gamow-Teller
operator connecting the ground state of the parent nucleus, which corresponds to the
absolute minimum in the DECs, with the excited states of the
daughter nucleus are calculated in the pnQRPA \cite{sarri1,sarri2,sarri3,sarri_odd}.
The GT strength is calculated on top of the HF+BCS basis with a spin-isospin residual 
interaction, which is reduced to a separable force. 
To illustrate the role of this residual interaction, 
Fig. \ref{fig_bgt} shows the energy distribution of the GT strength with and without
the spin-isospin separable force on the example of $^{260}$Sg, which is a 
representative nucleus in this mass region. The main effect observed is a reduction
of the GT strength that increases the $\beta$-decay half-life. 
Although the validity of this approach has still to be justified for SHN, the 
finite rank separable approximation \cite{vangiai_98} has been shown to describe
properly a large amount of nuclear properties and, in particular, charge-exchange 
excitations and spin-isospin properties in different mass regions \cite{severyukhin_2012}.

Other effects on the $\beta$-decay properties, such as tensor correlations 
and the coupling of one- and two-phonon configurations have been studied in Ref. 
\cite{sushenok_2020} for Cd isotopes, concluding that typical effects of around
30 \%  are expected in the half-lives, being smaller for heavier isotopes. 
Similarly, tensor interactions can have an effect on fission barriers and,
therefore, on the spontaneous fission half-lives \cite{tolokonnikov_2018}, 
but this would not significantly alter the ratio between $\alpha$- and $\beta$-decays.

In this work, only allowed GT transitions are considered, whereas first forbidden 
(FF) transitions are not included. FF transitions are expected to make relevant 
contributions only in cases where the dominant GT are suppressed and large Q-energy
windows provide large phase-space factors. When these conditions are met, although the 
nuclear matrix elements of FF transitions are orders of magnitude smaller than the GT 
ones, the phase factors that scale with the fifth power of the energy could make these
contributions relevant. However, in the SHN studied here these conditions are not
fulfilled and, therefore, FF contributions are not expected to play a significant 
role. This can be seen from existing calculations \cite{marketin_2016}, where 
FF contributions were shown to have little effect on the total $\beta$-decay 
half-lives of SHN.

This model has been used in the past to study GT strength distributions and 
$\beta$-decay half-lives in different regions of the nuclear chart.
These studies include medium-mass \cite{sarri_wp,sarri_rp1,sarri_rp2,sarri_rp3}
and heavy nuclei \cite{sarri_pb1,sarri_pb2,sarri_pb3} in the neutron-deficient 
side, as well as nuclei in the neutron-rich side 
\cite{sarripere1,sarripere2,sarripere3}, and $fp$-shell nuclei 
\cite{sarri_fp1,sarri_fp2,sarri_fp3}.
Spin magnetic dipole excitations, which are the $\Delta T_z=0$ counterparts of the 
GT excitations, were also studied within this approach in Refs. \cite{m1}.
The sensitivity of the GT strength distributions to the nuclear
deformation has been used to get information on the nuclear shape by comparing
theoretical results with $\beta$-decay data \cite{expnacher}.

The phase-space factors, $f^i$, contain two components, positron emission $f^{\beta^+}$ 
and electron capture $f^{EC}$. For a given nucleus, they are computed numerically 
for each energy value according to Ref. \cite{gove},

\begin{equation}
f^{\beta^+} (Z, W_0) = \int^{W_0}_1 p W (W_0 - W)^2 \lambda^+(Z,W) {\rm d}W\, ,
\label{phase}
\end{equation}
where the Fermi function $\lambda^+(Z,W)$ accounts for the Coulomb distortion
of the $\beta$ particle.

The phase factors for electron capture, $f^{EC}$, are given by

\begin{equation}
f^{EC}=\frac{\pi}{2} \sum_{x} q_x^2 g_x^2B_x \, ,
\end{equation}
where $x$ stands for the atomic sub-shell ($K$, $L$) from which the electron
is captured, $q$ is the energy of the neutrino, $g$ is the radial component of the
bound-state electron wave function at the nuclear surface, and $B$ stands for
various exchange and overlap corrections \cite{gove}.

\begin{figure}[th]
\includegraphics[width=65mm]{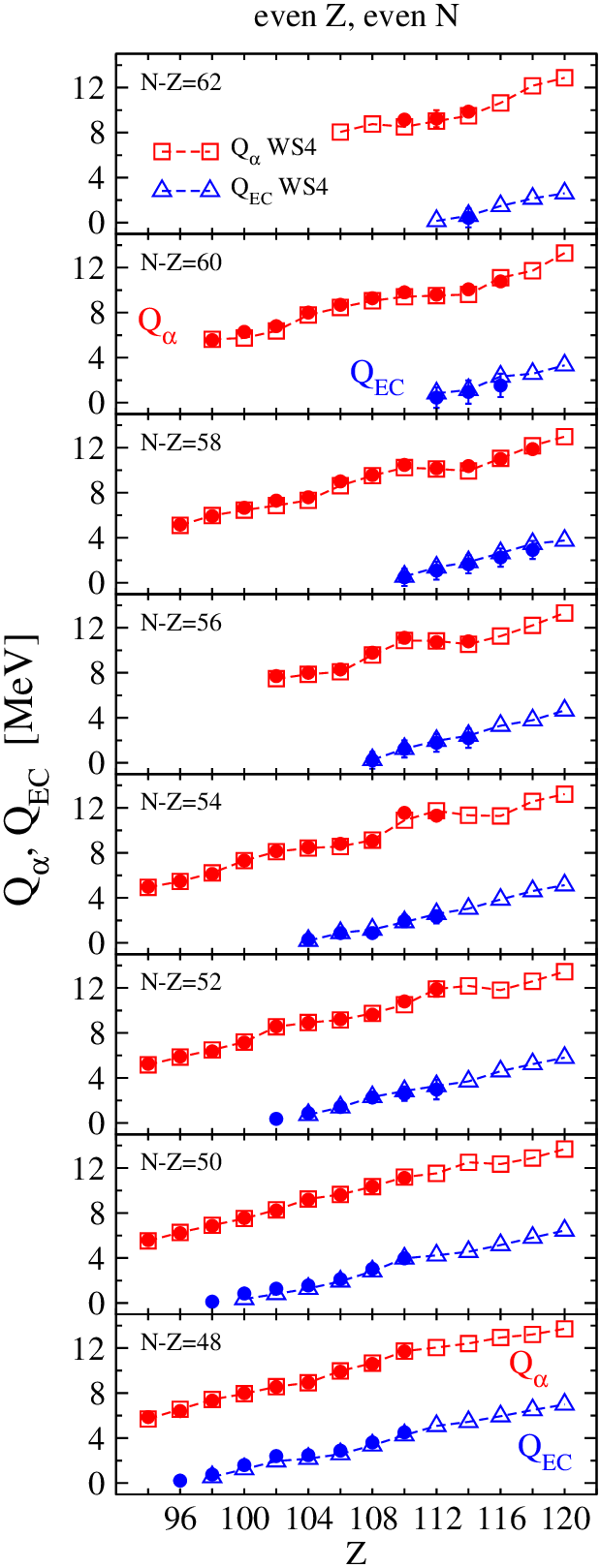}
\caption{$Q_{\alpha}$ (red) and $Q_{EC}$ (blue) energies for even-even nuclei involved 
in the $\alpha$-decay chains starting at $Z=120$ and characterized by different 
$N-Z$ values from 62 down to 48. Square and triangle open symbols correspond,
respectively, to $Q_{\alpha}$ and $Q_{EC}$ energies calculated with WS4 \cite{ws4}, 
whereas circles correspond to experimental values \cite{ame2020}.}
\label{fig1}
\end{figure}

\begin{figure}[th]
  \includegraphics[width=65mm]{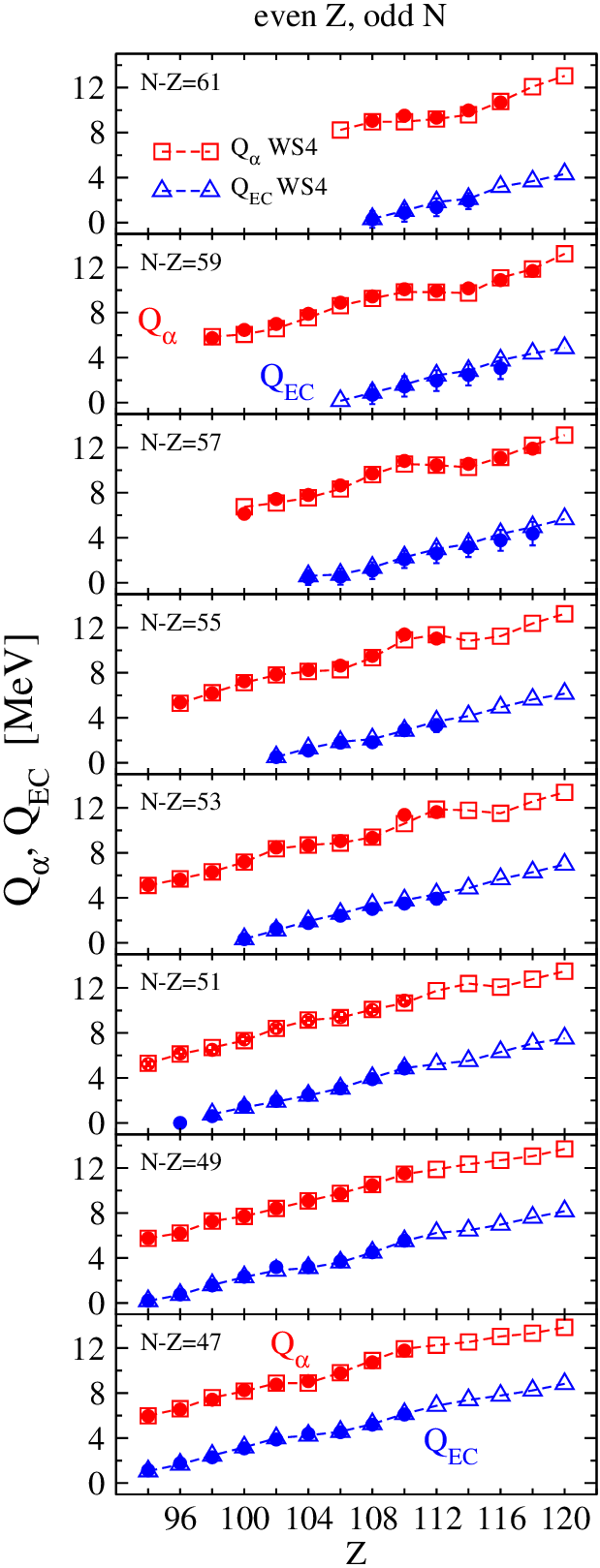}
\caption{Same as in Fig. \ref{fig1}, but for  $\alpha$-decay chains with
even $Z$ and odd $N$ nuclei.}
\label{fig2}
\end{figure}

\begin{figure}[th]
\includegraphics[width=65mm]{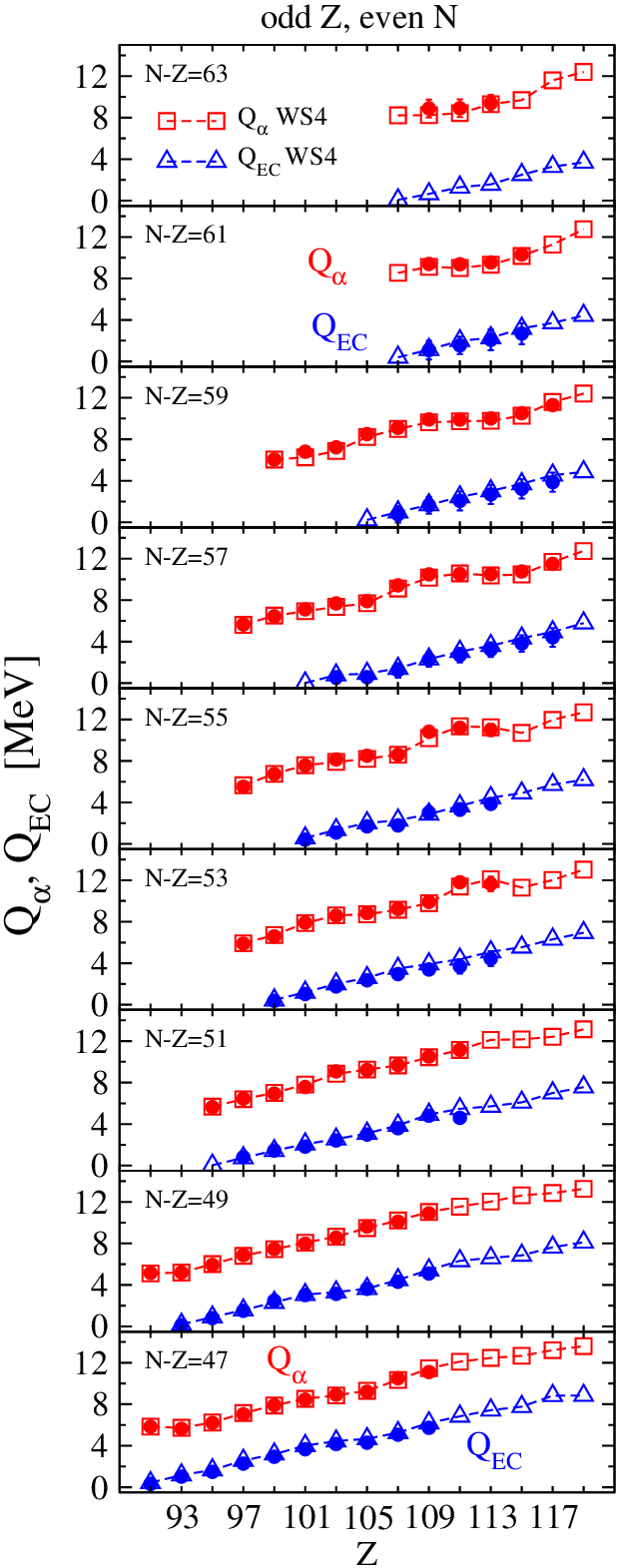}
\caption{Same as in Fig. \ref{fig1}, but for  $\alpha$-decay chains with odd $Z$ 
and even $N$ nuclei involved in the $\alpha$-decay chains starting at $Z=119$.}
\label{fig3}
\end{figure}

$\beta^+/EC$-decay half-lives depend also on the $Q$-energies.
These energies determine the maximum energy available in the process, as well as
the values of the phase factors, see Eq. (\ref{t12}).
They are given by

\begin{equation}
Q_{EC}=Q_{\beta^+} +2m_e= M(A,Z)-M(A,Z-1)+m_e \, , 
\label{qec}
\end{equation}
written in terms of the nuclear masses $M(A,Z)$ and the electron mass ($m_e$).

In the cases where the experimental masses are available \cite{ame2020}, these 
values are used to evaluate Eq. (\ref{qec}). But in cases where masses have
not been yet measured, theoretical predictions must be used.
A large number of mass formulas obtained from different approaches can be
found in the literature. They have been discussed for SHN in Refs. 
\cite{sarri_shn_prc,sarri_shn_jpg,sarri_shn_plb}.
In this work we use the masses (WS4+RBF) \cite{ws4} obtained from a
macroscopic-microscopic approach inspired by the Skyrme energy-density functional,
including a surface diffuseness correction for unstable nuclei and radial basis
function (RBF) corrections.
This mass formula has been shown to be very reliable to describe SHN \cite{wang15}.

\section{Results}

\begin{figure}[h]
  \includegraphics[width=65mm]{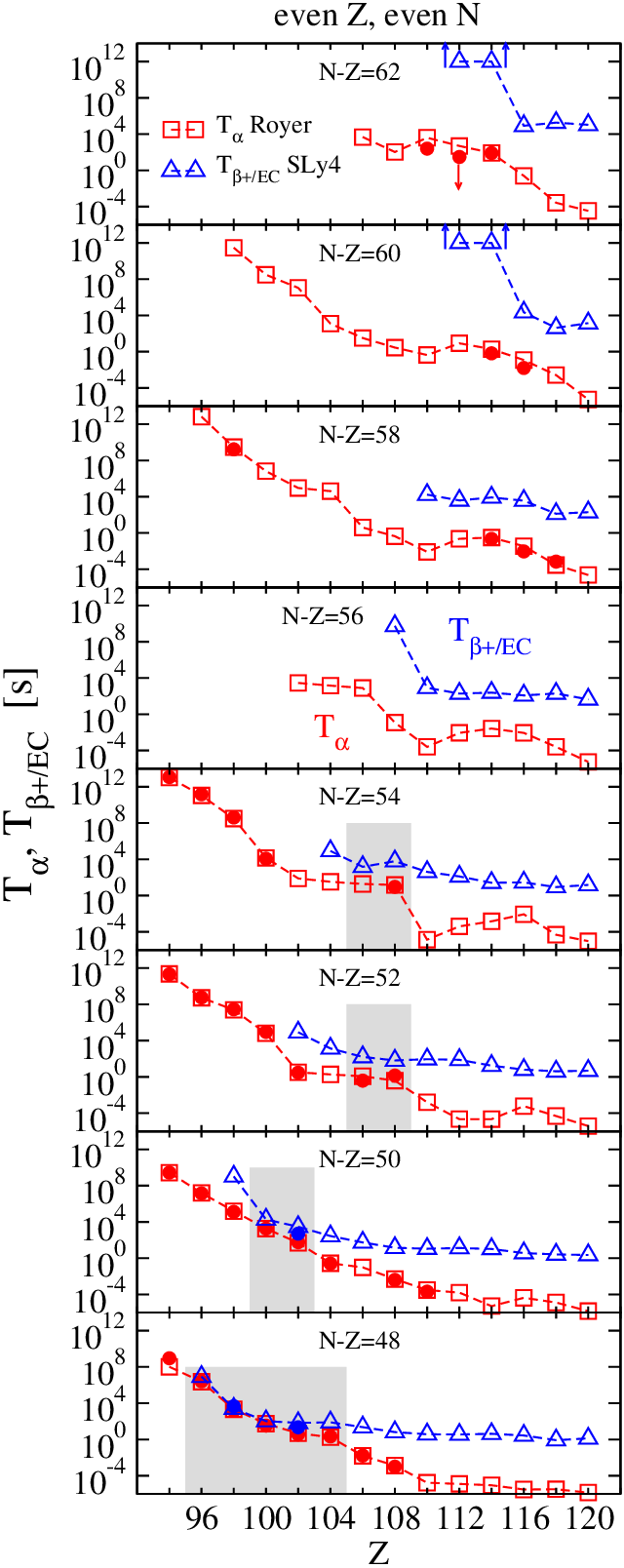}
\caption{Calculated half-lives for $\alpha$ (Royer, open squares) and $\beta^+/EC$
  (SLy4, open triangles) decays. Red and blue points correspond to experimental values
  \cite{nubase2020} for $\alpha$ and $\beta^+/EC$ decays, respectively. The gray 
shaded area highlights the cases where the competition between both decay modes is
significant. The set of nuclei considered corresponds to the same $\alpha$-decay 
members of Fig. \ref{fig1}.}
\label{fig4}
\end{figure}

\begin{figure}[h]
\includegraphics[width=65mm]{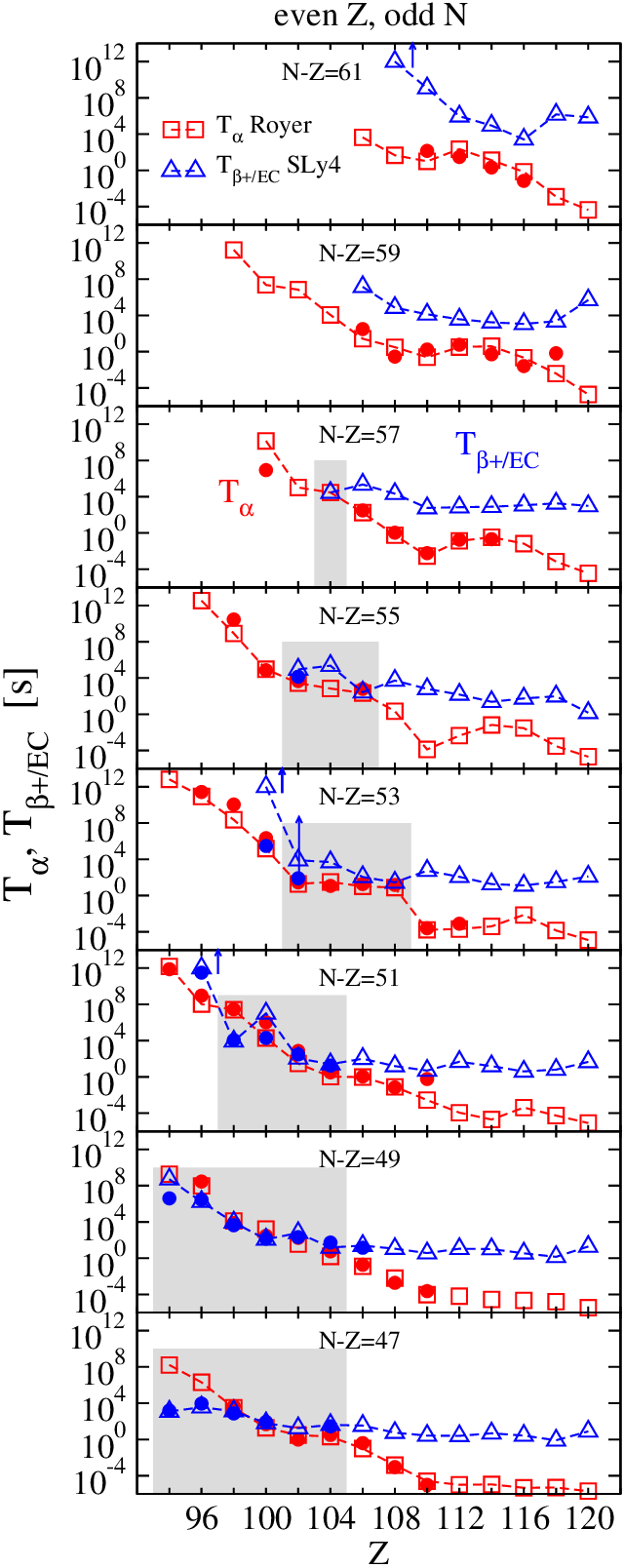}
\caption{Same as in Fig. \ref{fig4}, but for the set of nuclei of Fig. \ref{fig2}.}
\label{fig5}
\end{figure}

\begin{figure}[h]
\includegraphics[width=65mm]{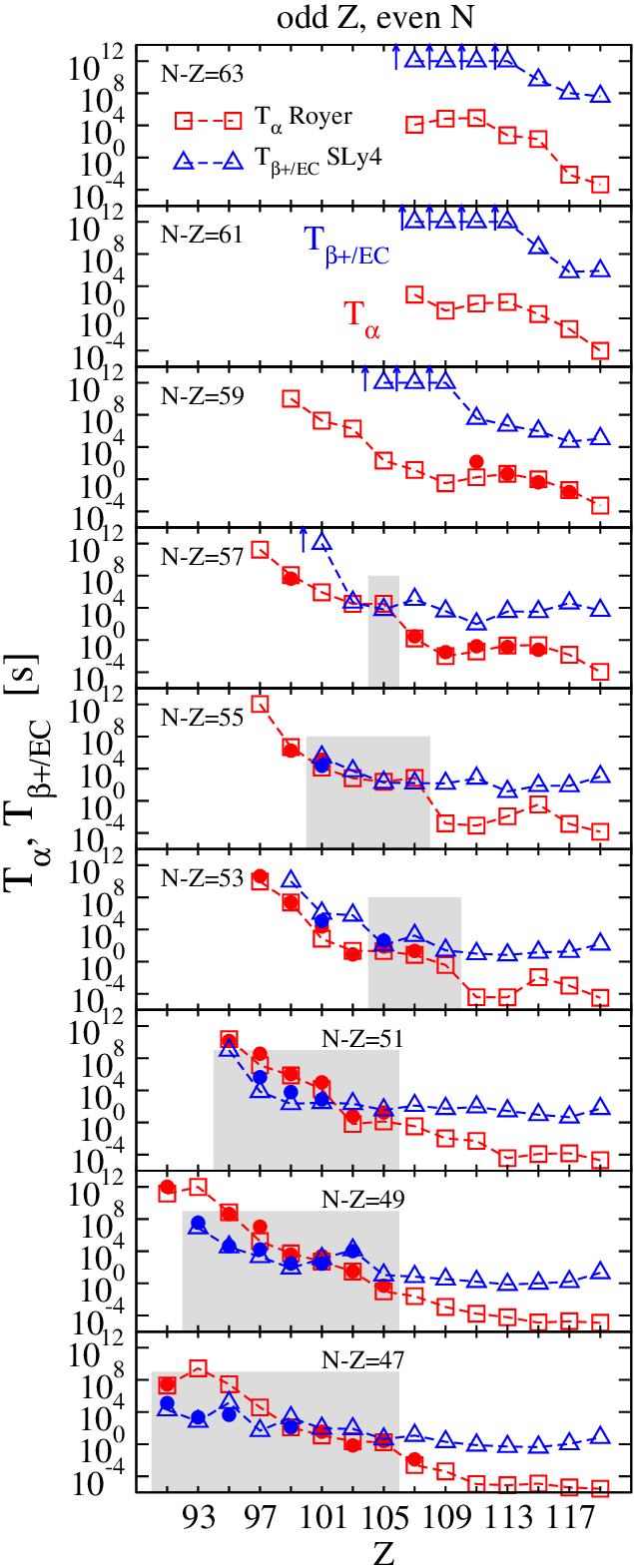}
\caption{Same as in Fig. \ref{fig4}, but for the set of nuclei of Fig. \ref{fig3}.}
\label{fig6}
\end{figure}

In this section we proceed to compare the half-lives for the $\alpha$ and
$\beta^+/EC$ decays for nuclei involved in $\alpha$-decay chains starting
at isotopes of $Z=119$ and $Z=120$. These $\alpha$-decay chains are characterized by
a fixed number of the difference $N-Z$. Thus, this value is used to label a given
chain. 

Comparison between $\alpha$- and $\beta^+/EC$-decay modes is critical to understand
the branching ratios and possible pathways of the original compound nucleus leading 
to stability. $\alpha$-decay half-lives are estimated in this work from 
phenomenological formulas, whereas $\beta^+/EC$-decay half-lives are calculated 
from the present microscopic calculations.
The experimental information of these decay branches in SHN is still
very limited. In the cases where experimental information on the total half-life,
as well as on the percentage of the corresponding mode intensity are available, 
the half-lives have been  extracted and plotted together with the calculations.

The half-lives depend on the $Q$-energies which are taken from experiment 
\cite{ame2020} or from Ref. \cite{ws4} when the masses are not measured.
Figs. \ref{fig1}-\ref{fig3} contain $Q_{\alpha}$ and $Q_{EC}$ energies obtained 
from experiment \cite{ame2020} and from WS4 \cite{ws4}.
The three figures correspond to the cases of nuclei involved in the $\alpha$ 
decays with even $Z$ and even $N$ (Fig. \ref{fig1}), with even $Z$ and odd $N$ 
(Fig. \ref{fig2}), and with odd $Z$ and even $N$ (Fig. \ref{fig3}).
Each panel is for a given $N-Z$ value that characterizes the $\alpha$-decay
chains. Only nuclei with $Q > 0$ are considered. As an example, the bottom panel 
in Fig. \ref{fig1} with $(N-Z)=48$ for even values  of $N$ and $Z$ corresponds 
to the following $\alpha$-decay chain:

\noindent
$^{288}$120 - $^{284}$Og - $^{280}$Lv -  $^{276}$Fl - $^{272}$Cn - $^{268}$Ds - $^{264}$Hs -
$^{260}$Sg - $^{256}$Rf - $^{252}$No - $^{248}$Fm - $^{244}$Cf - $^{240}$Cm - $^{236}$Pu.

Other panels with different values of $(N-Z)$ correspond to similar $\alpha$-decay
chains, but for different isotopes of a given $Z$.
The $\alpha$-decay chains in Fig. \ref{fig2} are similar to Fig. \ref{fig1}, but for 
odd values of $N$. Finally, as an example of the  $\alpha$-decay chains involving 
nuclei with odd $Z$ and even $N$ in  Fig. \ref{fig3}, the bottom panel 
corresponds to the following $\alpha$-decay chain:

\noindent
$^{285}$119 - $^{281}$Ts - $^{277}$Mc - $^{273}$Nh - $^{269}$Rg - $^{265}$Mt -
$^{261}$Bh - $^{257}$Db -  $^{253}$Lr -  $^{249}$Md -  $^{245}$Es -  $^{241}$Bk -
$^{237}$Am - $^{233}$Np - $^{229}$Pa,
and similarly for the other panels in this figure, but for different isotopes
of a given $Z$.

One can see that the experimental $Q$-values are well reproduced by the mass formula
WS4, justifying its use when no experimental information is available. It is worth 
noting the inflection point in $Q_{\alpha}$ at around $Z=110$ in the higher $N-Z$ 
values, showing a bump which is translated into a relative depression in the 
corresponding $T_{\alpha}$ values.

For a given $N-Z$ value in each panel, when $Z$ increases $N$ also increases by
the same amount and so do both energies $Q_{\alpha}$ and $Q_{EC}$, reflecting general 
properties of nuclear stability in terms of
the relative number of both types of nucleons.


\begin{table}[h]
  \centering
  \caption{Spin and parity $J^{\pi}$, quadrupole deformation  $\beta_2$, $\alpha$-decay 
half-lives $T_{\alpha}$, $\beta^+/EC$-decay half-lives $T_{\beta^+/EC}$, and their ratio 
$\cal{R}$ for members of different $\alpha$-decay chains with even numbers of protons 
and neutrons, characterized by $(N-Z)$. Only cases with ratios smaller than $10^3$ 
are considered. } 
{\begin{tabular}{ccccccc} \hline \hline \\
 N-Z &   Nucleus   & $J^{\pi}$ & $\beta_2$ & $T_{\alpha}$ [s] & $T_{\beta^+/EC}$ [s]
       & $\cal{R}$ \\ 
\\
 $54$ &  $^{266}$Sg  & $0^+$ & 0.244 & $ 1.90\times 10^{1}$ & $ 1.48\times 10^{3}$ 
               & $ 7.8\times 10^{1}$ \\
      &  $^{270}$Hs  & $0^+$ & 0.234 & $ 1.50\times 10^{1}$ & $ 6.12\times 10^{3}$ 
               & $ 4.1\times 10^{2}$ \\ \\
 $52$ &  $^{264}$Sg  & $0^+$ & 0.251 & $ 1.16\times 10^{0}$ & $ 1.57\times 10^{2}$ 
               & $ 1.4\times 10^{2}$ \\
      &  $^{268}$Hs  & $0^+$ & 0.239 & $ 3.84\times 10^{-1}$& $ 6.44\times 10^{1}$ 
               & $ 1.7\times 10^{2}$ \\ \\
 $50$ &  $^{250}$Fm  & $0^+$ & 0.276 & $ 1.85\times 10^{3}$ & $ 1.87\times 10^{4}$ 
               & $ 1.0\times 10^{1}$ \\
      &  $^{254}$No  & $0^+$ & 0.284 & $ 4.95\times 10^{1}$ & $ 2.89\times 10^{3}$ 
               & $ 5.8\times 10^{1}$ \\ \\
 $48$ &  $^{240}$Cm  & $0^+$ & 0.269 & $ 2.35\times 10^{6}$ & $ 8.50\times 10^{6}$ 
               & $ 3.6\times 10^{0}$ \\
      &  $^{244}$Cf  & $0^+$ & 0.276 & $ 2.16\times 10^{3}$ & $ 2.53\times 10^{3}$ 
               & $ 1.2\times 10^{0}$ \\
      &  $^{248}$Fm  & $0^+$ & 0.278 & $ 5.38\times 10^{1}$ & $ 1.06\times 10^{2}$  
               & $ 2.0\times 10^{0}$ \\
      &  $^{252}$No  & $0^+$ & 0.275 & $ 4.73\times 10^{0}$ & $ 6.64\times 10^{1}$ 
               & $ 1.4\times 10^{1}$ \\ 
      &  $^{256}$Rf  & $0^+$ & 0.271 & $ 1.78\times 10^{0}$ & $ 7.91\times 10^{1}$ 
               & $ 4.5\times 10^{1}$ \\ \\
\hline \hline
\label{table1}
\end{tabular}}
\end{table}


\begin{table}[h]
  \centering
  \caption{Same as in Table \ref{table1}, but for $\alpha$-decay chains with 
an even number of protons and an odd number of neutrons.} 
{\begin{tabular}{ccccccc} \hline \hline \\
 N-Z &   Nucleus   & $J^{\pi}$ & $\beta_2$ & $T_{\alpha}$ [s] & $T_{\beta^+/EC}$ [s]
       & $\cal{R}$ \\ 
\\
 $57$ &  $^{265}$Rf  & $9/2^+$ & 0.247 & $ 2.89\times 10^{4}$ & $ 2.80\times 10^{4}$ 
               & $ 9.7\times 10^{-1}$ \\ \\
 $55$ &  $^{259}$No  & $9/2^+$ & 0.259 & $ 2.74\times 10^{3}$ & $ 8.92\times 10^{4}$ 
               & $ 3.3\times 10^{1}$ \\
      &  $^{263}$Rf  & $3/2^+$ & 0.251 & $ 6.87\times 10^{2}$ & $ 2.30\times 10^{5}$ 
               & $ 3.4\times 10^{2}$ \\
      &  $^{267}$Sg  & $3/2^+$ & 0.244 & $ 2.18\times 10^{2}$ & $ 2.70\times 10^{2}$ 
               & $ 1.2\times 10^{0}$ \\ \\
 $53$ &  $^{257}$No  & $7/2^+$ & 0.277 & $ 1.90\times 10^{1}$ & $ 7.87\times 10^{3}$ 
               & $ 4.1\times 10^{2}$ \\
      &  $^{261}$Rf  & $3/2^+$ & 0.257 & $ 3.21\times 10^{1}$& $ 4.97\times 10^{3}$ 
               & $ 1.5\times 10^{2}$ \\
      &  $^{265}$Sg  & $9/2^+$ & 0.249 & $ 1.02\times 10^{1}$& $ 1.14\times 10^{2}$ 
               & $ 1.1\times 10^{1}$ \\
      &  $^{269}$Hs  & $9/2^+$ & 0.240 & $ 7.60\times 10^{0}$& $ 2.81\times 10^{1}$ 
               & $ 3.7\times 10^{0}$ \\ \\
 $51$ &  $^{247}$Cf  & $7/2^+$ & 0.276 & $ 2.50\times 10^{7}$ & $ 8.49\times 10^{3}$ 
               & $ 3.4\times 10^{-4}$ \\
      &  $^{251}$Fm  & $9/2^-$ & 0.285 & $ 1.93\times 10^{4}$ & $ 1.01\times 10^{7}$ 
               & $ 5.2\times 10^{2}$ \\ 
      &  $^{255}$No  & $7/2^+$ & 0.285 & $ 2.97\times 10^{1}$ & $ 1.10\times 10^{2}$ 
               & $ 3.7\times 10^{0}$ \\
      &  $^{259}$Rf  & $7/2^+$ & 0.264 & $ 1.06\times 10^{0}$ & $ 2.52\times 10^{1}$ 
               & $ 2.4\times 10^{1}$ \\ \\
 $49$ &  $^{237}$Pu  & $7/2^-$ & 0.261 & $ 1.82\times 10^{9}$ & $ 5.10\times 10^{8}$ 
               & $ 2.8\times 10^{-1}$ \\
      &  $^{241}$Cm  & $1/2^+$ & 0.269 & $ 9.57\times 10^{7}$ & $ 1.70\times 10^{6}$ 
               & $ 1.8\times 10^{-2}$ \\
      &  $^{245}$Cf  & $1/2^+$ & 0.274 & $ 1.25\times 10^{4}$ & $ 8.94\times 10^{3}$  
               & $ 7.2\times 10^{-1}$ \\
      &  $^{249}$Fm  & $7/2^+$ & 0.277 & $ 1.66\times 10^{3}$ & $ 1.16\times 10^{2}$  
               & $ 7.0\times 10^{-2}$ \\
      &  $^{253}$No  & $9/2^-$ & 0.285 & $ 3.53\times 10^{1}$ & $ 6.21\times 10^{2}$ 
               & $ 1.8\times 10^{1}$ \\
      &  $^{257}$Rf  & $1/2^+$ & 0.270 & $ 1.58\times 10^{0}$ & $ 1.50\times 10^{1}$ 
               & $ 9.5\times 10^{0}$ \\ \\
$47$  &  $^{235}$Pu  & $5/2^+$ & 0.253 & $ 1.62\times 10^{8}$ & $ 1.23\times 10^{3}$ 
               & $ 7.6\times 10^{-6}$ \\
      &  $^{239}$Cm  & $1/2^+$ & 0.267 & $ 2.02\times 10^{6}$ & $ 3.53\times 10^{3}$ 
               & $ 1.7\times 10^{-3}$ \\
      &  $^{243}$Cf  & $1/2^+$ & 0.273 & $ 3.10\times 10^{3}$ & $ 1.34\times 10^{3}$  
               & $ 4.3\times 10^{-1}$ \\
      &  $^{247}$Fm  & $7/2^+$ & 0.285 & $ 1.98\times 10^{1}$ & $ 6.36\times 10^{1}$  
               & $ 3.2\times 10^{0}$ \\
      &  $^{251}$No  & $7/2^+$ & 0.276 & $ 3.03\times 10^{0}$ & $ 1.81\times 10^{1}$  
               & $ 6.0\times 10^{0}$ \\
      &  $^{255}$Rf  & $9/2^-$ & 0.272 & $ 2.07\times 10^{0}$ & $ 4.12\times 10^{1}$ 
               & $ 2.0\times 10^{1}$ \\ \\
\hline \hline
\label{table2}
\end{tabular}}
\end{table}


\begin{table}[h]
  \centering
  \caption{Same as in Table \ref{table1}, but for $\alpha$-decay chains with an 
odd number of protons and an even number of neutrons.} 
{\begin{tabular}{ccccccc} \hline \hline \\
 N-Z &   Nucleus   & $J^{\pi}$ & $\beta_2$ & $T_{\alpha}$ [s] & $T_{\beta^+/EC}$ [s]
       & $\cal{R}$ \\ 
\\
 $57$ &  $^{263}$Lr  & $7/2^+$ & 0.245 & $ 3.10\times 10^{4}$ & $ 5.03\times 10^{4}$ 
               & $ 1.6\times 10^{0}$ \\
      &  $^{267}$Db  & $9/2^+$ & 0.239 & $ 3.20\times 10^{4}$ & $ 6.20\times 10^{3}$ 
               & $ 1.9\times 10^{-1}$ \\ \\
 $55$ &  $^{257}$Md  & $7/2^+$ & 0.278 & $ 1.37\times 10^{4}$ & $ 2.74\times 10^{5}$ 
               & $ 2.0\times 10^{1}$ \\
      &  $^{261}$Lr  & $7/2^+$ & 0.253 & $ 6.52\times 10^{2}$ & $ 5.41\times 10^{3}$ 
               & $ 8.3\times 10^{0}$ \\
      &  $^{265}$Db  & $9/2^+$ & 0.245 & $ 2.39\times 10^{2}$ & $ 1.92\times 10^{2}$ 
               & $ 8.0\times 10^{-1}$ \\
      &  $^{269}$Bh  & $9/2^+$ & 0.236 & $ 6.94\times 10^{2}$ & $ 1.58\times 10^{2}$ 
               & $ 2.3\times 10^{-1}$ \\ \\
 $53$ &  $^{263}$Db  & $9/2^+$ & 0.252 & $ 2.08\times 10^{1}$ & $ 1.22\times 10^{2}$ 
               & $ 5.9\times 10^{0}$ \\
      &  $^{267}$Bh  & $5/2^-$ & 0.242 & $ 6.85\times 10^{0}$ & $ 1.76\times 10^{3}$ 
               & $ 2.6\times 10^{2}$ \\
      &  $^{271}$Mt  & $9/2^+$ & 0.227 & $ 3.87\times 10^{-1}$ & $ 2.63\times 10^{1}$ 
               & $ 6.8\times 10^{1}$ \\ \\
 $51$ &  $^{241}$Am  & $5/2^+$ & 0.265 & $ 2.31\times 10^{10}$ & $ 9.50\times 10^{8}$ 
               & $ 4.1\times 10^{-2}$ \\
      &  $^{245}$Bk  & $5/2^+$ & 0.270 & $ 1.33\times 10^{7}$ & $ 6.95\times 10^{3}$ 
               & $ 5.2\times 10^{-4}$ \\ 
      &  $^{249}$Es  & $7/2^+$ & 0.272 & $ 7.12\times 10^{5}$ & $ 2.10\times 10^{2}$ 
               & $ 2.9\times 10^{-4}$ \\
      &  $^{253}$Md  & $7/2^+$ & 0.281 & $ 1.38\times 10^{4}$ & $ 2.97\times 10^{2}$ 
               & $ 2.2\times 10^{-2}$ \\
      &  $^{257}$Lr  & $7/2^+$ & 0.266 & $ 6.64\times 10^{-1}$ & $ 2.17\times 10^{2}$ 
               & $ 3.2\times 10^{2}$ \\
      &  $^{261}$Db  & $9/2^+$ & 0.260 & $ 1.36\times 10^{0}$ & $ 3.61\times 10^{1}$ 
               & $ 2.7\times 10^{1}$ \\ \\
 $49$ &  $^{235}$Np  & $3/2^+$ & 0.247 & $ 9.66\times 10^{11}$ & $ 7.90\times 10^{6}$ 
               & $ 8.2\times 10^{-6}$ \\
      &  $^{239}$Am  & $5/2^+$ & 0.260 & $ 6.85\times 10^{8}$ & $ 3.23\times 10^{4}$ 
               & $ 4.7\times 10^{-5}$ \\
      &  $^{243}$Bk  & $5/2^+$ & 0.268 & $ 1.78\times 10^{5}$ & $ 2.07\times 10^{3}$  
               & $ 1.2\times 10^{-2}$ \\
      &  $^{247}$Es  & $7/2^+$ & 0.273 & $ 5.32\times 10^{3}$ & $ 7.78\times 10^{1}$  
               & $ 1.5\times 10^{-2}$ \\
      &  $^{251}$Md  & $7/2^-$ & 0.271 & $ 5.13\times 10^{2}$ & $ 1.18\times 10^{3}$ 
               & $ 2.3\times 10^{0}$ \\ 
      &  $^{255}$Lr  & $7/2^-$ & 0.279 & $ 3.02\times 10^{1}$ & $ 1.37\times 10^{4}$ 
               & $ 4.5\times 10^{2}$ \\
      &  $^{259}$Db  & $9/2^+$ & 0.265 & $ 1.00\times 10^{-1}$ & $ 1.07\times 10^{1}$ 
               & $ 1.1\times 10^{2}$ \\ \\
$47$ &  $^{229}$Pa  & $5/2^+$ & 0.211 & $ 1.94\times 10^{7}$ & $ 1.87\times 10^{4}$ 
               & $ 1.0\times 10^{-3}$ \\
      &  $^{233}$Np  & $5/2^+$ & 0.236 & $ 2.67\times 10^{9}$ & $ 6.73\times 10^{2}$ 
               & $ 2.5\times 10^{-7}$ \\
      &  $^{237}$Am  & $5/2^-$ & 0.252 & $ 2.79\times 10^{7}$ & $ 1.45\times 10^{5}$  
               & $ 5.2\times 10^{-3}$ \\
     &  $^{241}$Bk  & $7/2^+$ & 0.264 & $ 3.78\times 10^{4}$ & $ 4.78\times 10^{1}$  
               & $ 1.3\times 10^{-3}$ \\
     &  $^{245}$Es  & $3/2^-$ & 0.271 & $ 1.25\times 10^{2}$ & $ 1.92\times 10^{3}$  
               & $ 1.5\times 10^{1}$ \\
     &  $^{249}$Md  & $7/2^-$ & 0.271 & $ 1.23\times 10^{1}$ & $ 9.11\times 10^{1}$ 
               & $ 7.4\times 10^{0}$ \\ 
     &  $^{253}$Lr  & $7/2^-$ & 0.272 & $ 2.25\times 10^{0}$ & $ 8.05\times 10^{1}$ 
               & $ 3.6\times 10^{1}$ \\ 
     &  $^{257}$Db  & $9/2^+$ & 0.269 & $ 1.73\times 10^{0}$ & $ 4.13\times 10^{0}$ 
               & $ 2.4\times 10^{0}$ \\ \\
\hline \hline
\label{table3}
\end{tabular}}
\end{table}

$\alpha$-decay half-lives have been calculated with a representative phenomenological 
formula by Royer \cite{royer_00} that produces very reasonable results in the cases 
studied previously \cite{sarri_shn_prc,sarri_shn_jpg,sarri_shn_plb}. Other formulas 
available in the literature produce similar results, which do not differ much from the 
Royer formula for the purpose of this work 
\cite{sarri_shn_prc,sarri_shn_jpg,sarri_shn_plb}.

These values are compared in Figs. \ref{fig4}-\ref{fig6} with the experimentally
extracted $\alpha$-decay half-lives \cite{nubase2020}, showing a good agreement
between them. The figures also contain the experimental and microscopically
calculated $\beta^+/EC$-decay half-lives, with good agreement where
they have been measured, supporting the reliability of the calculations.

The values of the half-lives
$T_{\beta^+/EC}$ in a given $\alpha$-decay chain characterized by $N-Z$,
exhibit a rather flat behavior in the higher $Z$  nuclei and increase gradually
in the lighter partners of the $\alpha$-decay chain as $Q_{EC}$ decreases
in the lighter nuclides of each $\alpha$-decay chain.

In most cases the decays are dominated by the $\alpha$ mode. This is especially
true for nuclei with higher $Z$ ($ Z=110-120$). However, one can find regions where 
the competition between both decay modes is important. In particular, in the 
$\alpha$-decay chains with $N-Z$ below 55 and with $Z$ below 108, this competition 
is relevant. These regions are highlighted with a gray area in the figures for 
half-lives.
Tables \ref{table1}-\ref{table3} contain a more detailed information in these
regions, including the spin-parity $J^\pi$ and the self-consistent quadrupole 
deformation $\beta_2$ of the ground states used in the calculations, the 
$\alpha$- and $\beta^+/EC$-decay half-lives, and their ratio 
${\cal{R}}=T_{\beta^+/EC}/T_{\alpha}$. Only nuclei with ratios $\cal{R}$ smaller than
$10^3$ are included. Similarly to the criterion used to group the different
$\alpha$-decay chains, Table \ref{table1} contains the cases with even numbers
of protons and neutrons, Table \ref{table2} is for $\alpha$-decay chains with even
$Z$ and odd $N$, whereas  Table \ref{table3} collects the results for chains with 
odd $Z$ and even $N$.

Actually, there are cases where the calculated weak decay is faster than the 
$\alpha$ decay. This is found specifically in the cases that follows:
$^{247}$Cf with $(N-Z)=51$; $^{237}$Pu, $^{241}$Cm, $^{245}$Cf, and $^{249}$Fm 
with $(N-Z)=49$; and $^{235}$Pu, $^{239}$Cm, and $^{243}$Cf with $(N-Z)=47$ in 
the case of even $Z$ and odd $N$ nuclei. 
In the case of odd $Z$ and even $N$ nuclei it is found for
$^{267}$Db with $(N-Z)=57$,
$^{265}$Db and $^{269}$Bh with $(N-Z)=55$, for $^{241}$Am, $^{245}$Bk, $^{249}$Es, 
and $^{253}$Md with $(N-Z)=51$, for $^{235}$Np, $^{239}$Am, $^{243}$Bk, 
and $^{247}$Es with $(N-Z)=49$, and for $^{229}$Pa, $^{233}$Np, $^{237}$Am, 
and $^{241}$Bk with $(N-Z)=47$.

\section{Conclusions}

In this work the weak-decay half-lives of the nuclides involved in the 
$\alpha$-decay chains that follow the synthesis of SHN are studied. 
The method of calculation is based on a self-consistent Skyrme HF+BCS 
approach. The $\beta^+/EC$-decay half-lives are compared with those of 
$\alpha$ decay obtained from phenomenological formulas, as well as with 
the available experimental data.

Although $\alpha$ decay and SF are the dominant decay modes in most of the 
SHN studied in this work, it is found that weak decays increase progressively
their relative importance when one moves towards lower values of the atomic 
number ($Z<108$) in a given chain, as well as when one moves towards 
$\alpha$-decay chains with lower values of the difference between neutrons 
and protons ($N-Z<55$). In fact, the present calculations have identified 
interesting regions where the $\beta^+/EC$ decay is faster than the $\alpha$ 
decays.

Improved treatments of the residual interaction and inclusion of tensor forces 
and first forbidden transitions might have an effect on the $\beta$-decay 
half-lives studied here and it would certainly be worth exploring their
impact in the future. However, according to previous studies of these effects 
in other mass regions, it could be safely established that the conclusions of
this work regarding the competition between $\alpha$- and $\beta$-decay 
modes will not be significantly altered.

These findings could be used as an experimental guide in searching for weak-decay
modes in SHN, pointing to the most favorable cases worth exploring experimentally,
where weak decays could be more easily observed.

\begin{acknowledgments}
  
  This work was supported by Ministerio de Ciencia e Innovaci\'on  MCI/AEI/FEDER,UE
  (Spain) under Contract No. PGC2018-093636-B-I00.

\end{acknowledgments}

\end{document}